%% file: main.tex
\documentclass[runningheads]{llncs}
\usepackage{graphicx}
\usepackage{diagbox}
\usepackage{makecell, booktabs}
\usepackage{comment}
\usepackage{lscape}

\usepackage[normalem]{ulem}
\useunder{\uline}{\ul}{}
\usepackage{tabularx}
\usepackage{ragged2e}
\usepackage{enumitem}
\usepackage{wrapfig}
\usepackage{multirow}
\usepackage[hyphens]{url}
\usepackage{csquotes}

\usepackage{tikz}
\usetikzlibrary{calc}
\usetikzlibrary{matrix}
\usetikzlibrary{positioning}
\usetikzlibrary{shapes.geometric}
\usepackage{booktabs}%

\begin{document}
\title{A framework to evaluate the viability of robotic process automation for business process activities}
\titlerunning{A framework for RPA viability assessment}
\author{
Christian Wellmann\inst{1} \and
Matthias Stierle\inst{1}
\and
Sebastian Dunzer\inst{1}
\and
Martin Matzner\inst{1}
}
\authorrunning{C. Wellmann et al.}
\institute{Friedrich-Alexander-Universit\"at Erlangen-N\"urnberg, Institute of Information Systems, N\"urnberg, Germany \\
\email{\{christian.wellmann, matthias.stierle,sebastian.dunzer, martin.matzner\}@fau.de}\\
\url{http://is.rw.fau.eu}}
\maketitle              %
\begin{abstract}
Robotic process automation (RPA) is a technology for centralized automation of business processes. 
RPA automates user interaction with graphical user interfaces, 
whereby it promises efficiency gains and a reduction of human negligence during process execution.
To harness these benefits, organizations face the challenge of classifying process activities as viable automation candidates for RPA.
Therefore, this work aims to support practitioners in evaluating RPA automation candidates. 
We design a framework that consists of thirteen criteria grouped into five perspectives which offer different evaluation aspects.
These criteria leverage a profound understanding of the process step.
We demonstrate and evaluate the framework by applying it to a real-life data set.
\end{abstract}

This is an accepted manuscript for the \emph{RPA Forum} at the \emph{Int. Conference on Business Process Management (BPM 2020)}. The final authenticated version is available online at \url{https://doi.org/10.1007/978-3-030-58779-6_14}.

\input{sections/01_intro}
\input{sections/02_literature}
\input{sections/04_artefact}

\input{sections/05_evaluation}
\input{sections/06_discussion}

\subsubsection*{Acknowledgments}
This project is funded by the German Federal Ministry of Education and Research (BMBF) within the framework programme \textit{Software Campus} (https://softwarecampus.de) under the number 01IS17045.

\bibliographystyle{splncs04}
\bibliography{mybibliography}

\end{document}

%% file: sections/01_intro.tex
\section{Introduction}

The state of technology is continuously advancing, resulting in shorter intervals to scrutinize whether tasks can be automated or rely on human execution \cite{van2018robotic}.
The recent rise of robotic process automation (RPA) challenges this status quo once more and further blurs the boundaries of human computer interaction  \cite{leshob2018towards}.
RPA automates repetitive and monotonous tasks by configuring software robots to mimic the actions of the user on the presentation layer \cite{aguirre2017automation}.
Organizations are hoping for RPA to lead to an increase in time for employees to focus on value-adding activities and to cut costs \cite{kaya2019impact} through eliminating time spent interacting with information systems and data transfer \cite{syed2020robotic}.
Furthermore, companies expect RPA to improve the quality of their work, eliminate human negligence and increase reaction time around the clock \cite{fernandez2018impacts}.
Primarily driven by changing market dynamics and global competition, companies are forced to cut costs through the implementation of new technologies like RPA, especially when they promise a quick and high return on investment \cite{van2018robotic}.
\par
While the benefits for organizations in applying RPA seem evident, the question remains as to why there are currently only few success stories of RPA adoption. 
One of the biggest challenges identified for a successful RPA implementation is the selection of suitable processes or process activities for RPA \cite{Hindel.2020,syed2020robotic,van2018robotic}.
The methods available to date mostly offer high-level decision-making support with the focus set on profitability rather than assessing the RPA viability of processes or tasks~\cite{beetz2019robotic,leshob2018towards,wanner2019process}.

The objective of this work is to offer practitioners a process characteristic evaluation framework including a set of criteria and exemplary evaluation metrics. %
To understand the parameters of RPA, the following research question needs to be answered:
\par
\textit{What are the characteristics of a process activity, or a set of process activities, that facilitate viable robotic process automation?}
\par
By answering the question, this work contributes to broadening the understanding in the selection of process activities for RPA.
Furthermore, it serves as a basis for the creation of a framework that examines the process activity from different perspectives for its suitability for RPA.
In addition, the application of the framework highlights challenges when assessing the criteria and opens up new research opportunities. %
\par
This study is structured as follows: In Section 2, the term robotic process automation is defined and the results of a literature review are presented as a concept matrix.
Further, the existing methods for process or process activity selection are compared to derive the similarities and differences.
In Section 3, the process characteristic evaluation framework is presented.
Section 4 outlines the evaluation approach, the data set and the pre-processing of the process before the framework is applied and validated.
The contributions, limitations and future research are summarized in Section 5.

%% file: sections/02_literature.tex
\section{Background}

\subsection{Robotic Process Automation (RPA)}
\label{chap:2.1}
While the interest in RPA is still steadily increasing \cite{santos2019toward}, there is no well accepted definition found in literature.
Despite the arguable lack of definition, certain characteristics describing the term Robotic Process Automation are found throughout the literature.
\par
RPA incorporates different tools and methodologies \cite{fernandez2018impacts,mendling2018machine,ratia2018robotic,van2018robotic} aiming to automate repetitive and structured service tasks that were previously performed by humans \cite{aguirre2017automation,lacity2017new,van2018robotic}.
This is achieved by the application of software algorithms known as {software robots} or \emph{bots}, which are imitating the execution flow of humans on the front-end \cite{aguirre2017automation,geyer2018process,Ivancic.2019,moffitt2018robotic,penttinen2018choose,van2018robotic}.
Just as a human user, robots can interact with the user interface through mouse clicks, key board interactions and interpretation of text and graphics~\cite{penttinen2018choose}, as well as log into multiple applications to extract, process and enter structured or semi-structured data from different sources~\cite{wanner2019process}.
RPA usually does not require defined interfaces as the software sits on top of information systems and accesses applications only through the presentation layer~\cite{aguirre2017automation,willcocks2016service}, thus the back-end systems remain unchanged \cite{lacity2015robotic,van2018robotic}.
As a result the robots perform activities in a non-invasive manner \cite{Ivancic.2019} without the need of application programming interfaces (API) to transfer and process data \cite{wanner2019process}.

Depending on the configuration approach for software robots, little to no programming knowledge is required to implement and manage the orchestration and execution of the robots often referred to as \emph{low-code development} \cite{Hindel.2020,Ivancic.2019,lacity2015robotic,penttinen2018choose,osmundsen.2019}.
Although RPA typically favors less complex and cognitive tasks, advances in machine learning can extend the range of RPA application in the future \cite{anagnoste2017robotic,wanner2019process}. 
\par
In this work, we define RPA as an automation technology which performs work on the presentation layer, can be set up by a business user, and is managed on a centralized platform.

\subsection{Process characteristics of automatable activities}
\label{chap:2.2}

In order to develop the framework, the question - \textit{What are the characteristics of a process activity, or a set of process activities, that are suitable for robotic process automation?} - must first be answered.
To obtain a comprehensive list of potential process characteristic evaluation criteria, a literature review following the guidelines proposed by \cite{webster.2002} is conducted.
For an exhaustive review, sources are searched for in the databases Scopus, Google Scholar, and IEEE Xplore Digital library.
The identified criteria are then compiled, checked for redundancy and listed in a concept matrix (Table \ref{fig:conceptmatrix}) that relates the criteria with the source articles and visualizes the acceptance and relevance through the number of mentions. In particular, we used the criteria presented by Wanner et al. \cite{wanner2019process} as a starting point and extended the list through several iterations.

Ideal candidate processes for automation must be standardized
\cite{aguirre2017automation,anagnoste2017robotic,asatiani2016turning,beetz2019robotic,cooper2019robotic,fernandez2018impacts,fung2014criteria,geyer2018process,Hallikainen.2018,Hindel.2020,Hofmann-Rpa,huang2019applying,Ivancic.2019,jimenez.2019,kokina2019early,lacity2017new,lacity2015robotic,leshob2018towards,moffitt2018robotic,osmundsen.2019,penttinen2018choose,romao2019robotic,santos2019toward,seasongood2016not,sonmez2019conceptual,syed2020robotic,wanner2019process,willcocks2015robotic,yatskiv2019improved,zhang2019intelligent}. 
Therefore, the process or task needs to be strictly defined and structured 
\cite{anagnoste2017robotic,huang2019applying,jimenez.2019,moffitt2018robotic,wanner2019process,yatskiv2019improved}.
A high degree of standardization before automation is necessary to result in a low amount of process variations and outcomes
\cite{wanner2019process}.
No or low subjective judgment or interpretation skills
\cite{cooper2019robotic,jimenez.2019,syed2020robotic,yatskiv2019improved}
are required for decision making as the process follows a rule-based flow
\cite{beetz2019robotic,cooper2019robotic,Hindel.2020,Hofmann-Rpa,Ivancic.2019,kokina2019early,moffitt2018robotic,penttinen2018choose,santos2019toward,syed2020robotic,wanner2019process,yatskiv2019improved,zhang2019intelligent}.
Well-suited tasks for standardized processes are also mentioned to be mundane, simple and monotonous
\cite{cooper2019robotic,fernandez2018impacts,sonmez2019conceptual}. 

In combination with a high degree of standardization, the execution frequency of a process or task has a big impact on the automation potential.
In favor of RPA suitability, tasks need to be performed repetitively and in high transaction volumes 
\cite{cooper2019robotic,fernandez2018impacts,fung2014criteria,Hindel.2020,Hofmann-Rpa,huang2019applying,Ivancic.2019,jimenez.2019,kokina2019early,leshob2018towards,moffitt2018robotic,osmundsen.2019,penttinen2018choose,santos2019toward,sonmez2019conceptual,syed2020robotic,wanner2019process,yatskiv2019improved}.
Besides the volume of transactions it is mentioned that the transaction of a substantial amount of data implies an aptitude for RPA \cite{yatskiv2019improved}. 

Furthermore, the maturity of a process is an indicator as to whether it fulfills fundamental requirements for an automation effort. Maturity describes the frequency of changes to the logical execution flow of the process
\cite{beetz2019robotic,syed2020robotic}
and further, that the process and its tasks are specified, predictable, stable and measurable
\cite{Hindel.2020,huang2019applying,leshob2018towards,santos2019toward,syed2020robotic}.
Contrary to standardization, the failure rate describes the amount of deviations from the defined process flow.
Candidate processes suited for RPA show little or no amount of exceptions when tasks are being executed
\cite{beetz2019robotic,fung2014criteria,jimenez.2019,syed2020robotic,wanner2019process}
and do not require human intervention.
Additionally, the ratio of process tasks that undergo an unusual process flow or inhibit the structured flow to completion is limited or zero \cite{beetz2019robotic,wanner2019process}. 

\begin{table}[ht]
    \centering\settowidth\rotheadsize{Proneness to Human Error/}
    \resizebox{\textwidth}{!}{
    \input{figures/conceptmatrix}

    }
    \caption{Concept matrix with dimensions \cite{webster.2002}}
    \label{fig:conceptmatrix}
\end{table}

With the objective to further minimize the exceptions, stability of the systems in use and the process outcome is crucial.
For an execution following the predefined rules, the stability of user interfaces and the interaction between different systems is essential
\cite{penttinen2018choose}.
Ideal candidate tasks for RPA have as a result a limited number of exceptions and high predictability of their outcomes to avoid uncertainties and disruptions
\cite{fung2014criteria,wanner2019process}.   

The speed of tasks that require access and interaction with multiple systems can be increased immensely (e.g. data entry between systems).
In 17 out of 21 examined papers, tasks including the access to different systems are mentioned to be suitable for RPA (see Table \ref{fig:conceptmatrix}).
Whenever multiple systems need to be accessed by a user, the manual effort is high and also reflected by the time consumption for this task.
A software robot can work within the different systems flawlessly and execute the tasks more rapidly, enabling not only the extraction of information but also the triggering of events, when a task is completed 
\cite{anagnoste2017robotic,fernandez2018impacts,Ivancic.2019,jimenez.2019,osmundsen.2019}. 

In order for process activities to be performed between multiple systems, the data needs to be in a structured and digital form.
When data is structured
\cite{beetz2019robotic,cooper2019robotic,huang2019applying,kokina2019early,moffitt2018robotic,osmundsen.2019,santos2019toward,syed2020robotic,wanner2019process,zhang2019intelligent}, the software robot can then successfully interpret the given input and follow the execution flow of the process activities.

Apart from process and process activity characteristics, literature mentions that proneness to human errors is also an indicator for RPA potential.
This assumption is based on the fact, that with increasing volume of tasks, humans will more likely cause exceptions by false entry or incorrect data manipulation than a program would
\cite{fernandez2018impacts,Hindel.2020,Ivancic.2019,jimenez.2019,santos2019toward,yatskiv2019improved}.

Moreover, a process or task can be judged by its impact or value to the business.
This is where literature does not provide a clear outline due to the small amount of mentions. While some argue that automation potential exists for processes with a low degree of business value~\cite{fernandez2018impacts}, others state that processes with a low execution frequency but a high business value are suitable candidates for automation~\cite{fung2014criteria,yatskiv2019improved}.

Focusing on the voluminous and repetitive processes, the number of users involved in the execution reflect another perspective on RPA suitability.
Kokina and Blanchet \cite{kokina2019early} indicate potential benefits where several people are performing the same processes, when these are repetitive and require no or low subjective judgment.
A different perspective highlights the handovers of work between different stakeholder across departments as a factor to consider~\cite{wanner2019process}.

Last, the execution time of a process is a criteria to assess the suitability of processes for RPA~\cite{wanner2019process}. Decreasing the time spent with repetitive and highly transactional jobs, increases time for employees to focus on more value-adding tasks \cite{anagnoste2017robotic}.

%% file: figures/conceptmatrix.tex
\begin{tabular}{|p{3.5cm}|>{\centering\arraybackslash}p{0.75cm}|>{\centering\arraybackslash}p{0.75cm}|>{\centering\arraybackslash}p{0.75cm}|>{\centering\arraybackslash}p{0.75cm}|>{\centering\arraybackslash}p{0.75cm}|>{\centering\arraybackslash}p{0.75cm}|>{\centering\arraybackslash}p{0.75cm}|>{\centering\arraybackslash}p{0.75cm}|>{\centering\arraybackslash}p{0.75cm}|>{\centering\arraybackslash}p{0.75cm}|>{\centering\arraybackslash}p{0.75cm}|}																			
\toprule
\hline%
	
\diagbox[height=1.05\rotheadsize,innerwidth=3.5cm]{\raisebox{3ex}{\textbf{Articles}}}{\raisebox{-4ex}{\textbf{Characteristics}}}	& \rotatebox[origin=c]{90}{Standardization}	& \rotatebox[origin=c]{90}{Frequency}	& \rotatebox[origin=c]{90}{Number of systems}	& \rotatebox[origin=c]{90}{Structuredness of data}	&  \rotatebox[origin=c]{90}{Maturity}	& \rotatebox[origin=c]{90}{Proneness to human error}	& \rotatebox[origin=c]{90}{Failure rate} & \rotatebox[origin=c]{90}{Stability}	 & \rotatebox[origin=c]{90}{Value}	& \rotatebox[origin=c]{90}{Handover of work}   & \rotatebox[origin=c]{90}{Execution time}
\\ \hline	

\cite{fung2014criteria}	&	&\textbullet	&\textbullet 	&	&	& 	&\textbullet	&\textbullet	&\textbullet 	&	&	\\ \hline

\cite{anagnoste2017robotic}	&\textbullet	&	&\textbullet 	& 	&	& 	&	&	& &	&			\\ \hline

\cite{fernandez2018impacts}	&\textbullet	&\textbullet	&\textbullet 	& 	&	&\textbullet	&	&	& 	&	&			\\ \hline

\cite{leshob2018towards}	&\textbullet	&\textbullet	& 	& 	&\textbullet	&	&	&	& 		&	&		\\ \hline

\cite{moffitt2018robotic}	&\textbullet	&\textbullet	& 	&\textbullet 	&\textbullet	& 	&	&	& 	&	&			\\ \hline

\cite{penttinen2018choose}	&\textbullet	&\textbullet	&\textbullet 	& 	&	& 	&	&\textbullet	& \textbullet	&	&			\\ \hline

\cite{beetz2019robotic}	&\textbullet	&	& 	& \textbullet	&\textbullet	& 	&\textbullet	&	& 	&	&			\\ \hline

\cite{cooper2019robotic}	&\textbullet	&\textbullet	&\textbullet	&\textbullet	&	&  &	&	& 	&	&			\\ \hline

\cite{Hofmann-Rpa}	&\textbullet	&\textbullet	&\textbullet 	& 	&	& 	&	&	& &	&				\\ \hline

\cite{huang2019applying}	&\textbullet	&\textbullet	& \textbullet	&\textbullet 	&\textbullet	& 	&	&	& 	&	&			\\ \hline

\cite{Ivancic.2019}	&\textbullet	&\textbullet	&\textbullet 	& 	&	& \textbullet	&	&	& 		&	&		\\ \hline

\cite{jimenez.2019}	&\textbullet	&\textbullet	&\textbullet  	& &	& \textbullet	&\textbullet		&  &	&	&			\\ \hline

\cite{kokina2019early}	&\textbullet	&\textbullet	&\textbullet 	&\textbullet 	&	& 	&	&	& &\textbullet &		\\ \hline

\cite{osmundsen.2019}	&\textbullet	&\textbullet	&\textbullet 	&\textbullet 	&	& 	&	&	& 		&	&		\\ \hline

\cite{santos2019toward}	&\textbullet	&\textbullet	& \textbullet	&\textbullet 	&\textbullet	& \textbullet	&	&\textbullet	& 	&	&			\\ \hline

\cite{sonmez2019conceptual}	&\textbullet	&\textbullet	&\textbullet 	& 	&	& 	&	&	& 			&	&	\\ \hline

\cite{wanner2019process}	&\textbullet	&\textbullet	& \textbullet	& \textbullet	&\textbullet	&\textbullet 	&\textbullet	&\textbullet	& 	&\textbullet &\textbullet			\\ \hline

\cite{yatskiv2019improved}	&\textbullet	&\textbullet	&\textbullet	& 	&	& \textbullet	&	&	& \textbullet		&	&		\\ \hline

\cite{zhang2019intelligent}	&\textbullet	&	&\textbullet	& 	&	& 	&	&	& &	&				\\ \hline

\cite{Hindel.2020}	&\textbullet	&\textbullet	& 	& 	&\textbullet	&\textbullet 	&	&	& 	&	&			\\ \hline

\cite{syed2020robotic}	&\textbullet	&\textbullet	& \textbullet	& \textbullet	&\textbullet	& 	&\textbullet	&	& 	&	&			\\ \hline

Total	&20	&18	&17	&9	&8	&7	&5	&4	&3 	&2	&1			\\ \hline

\toprule\hline																			
\end{tabular}

%% file: sections/04_artefact.tex
\section{Process characteristics evaluation framework}
To support practitioners in evaluating the viability of RPA for process activities, we summarized the findings of our literature review in a framework.

Table \ref{tab:PCEF} visualizes the process characteristics evaluation framework (PCEF).
We present five perspectives -- task, time, data, system, and human -- that contain several characteristics that analysts can use to evaluate a process accordingly. We present examples for evaluation of the criteria but the list is certainly not exhaustive. 
We decided to exclude \textit{value} as a criteria from the framework as it is implicitly covered by other criteria such as frequency and urgency.

\begin{table}[!htb]
    \centering
    \begin{tabular}{p{.17\textwidth} p{.25\textwidth} p{.6\textwidth}}
    \toprule
    \textbf{Perspective} &
      \textbf{Criteria} &
      \textbf{Exemplary Evaluation} \\\midrule
    \multirow{13}*{Task}
       & Standardization
       &\vspace{-0.6cm}
       \begin{itemize}[leftmargin=0pt]
          \item[] Number of different activities
          \item[] Number of variations to execution flow in business
        \end{itemize}
           \\ 
     &
      Maturity &\vspace{-0.6cm}
      \begin{itemize}[leftmargin=0pt]
          \item[] Number of deviation cases over time
          \item[] Ratio of deviation cases over time
      \end{itemize} \\
           &
      &\vspace{-0.6cm}
      \begin{itemize}[leftmargin=0pt]
          \item[] Number of deviation cases over time
          \item[] Ratio of deviation cases over time
      \end{itemize} \\
     &
      Determinism & \vspace{-0.6cm}
      \begin{itemize}[leftmargin=0pt]
          \item[] Number of manual interactions
          \item[] Time to solve manual interaction
      \end{itemize} \\
     &
      Failure rate & \vspace{-0.6cm}
      \begin{itemize}[leftmargin=0pt]
          \item[] Number of unsuccessful terminations
          \item[] Number of manual interactions
          \item[] Number of rework loops
      \end{itemize} \\ \midrule
    \multirow{6}*{Time} &
      Frequency & \vspace{-0.6cm}
      \begin{itemize}[leftmargin=0pt]
          \item[] Number of executions
      \end{itemize} \\ 
     &
      Duration & \vspace{-0.6cm}
      \begin{itemize}[leftmargin=0pt]
          \item[] Average time to task completion
      \end{itemize} \\ 
     &
      Urgency & \vspace{-0.6cm}
      \begin{itemize}[leftmargin=0pt]
          \item[] Average reaction time
      \end{itemize}\\ \midrule
    \multirow{1}*{Data} &
      Structuredness & \vspace{-0.6cm}
       \begin{itemize}[leftmargin=0pt]
           \item[] Consistent use of data objects
       \end{itemize}\\ \midrule
    \multirow{8}*{System} &
      Interfaces & \vspace{-0.6cm}
      \begin{itemize}[leftmargin=0pt]
          \item[] Number of execution steps
          \item[] Time spent on application interface
      \end{itemize} \\ 
     &
      Stability &\vspace{-0.6cm}
      \begin{itemize}[leftmargin=0pt]
          \item[] Number of exceptions
      \end{itemize} \\ 
     &
      Number of systems &\vspace{-0.6cm}
      \begin{itemize}[leftmargin=0pt]
          \item[] Number of systems involved \newline
                (e.g. CRM, ERP)
      \end{itemize} \\\midrule
    \multirow{6}*{Human} &
      Resources &
      \begin{itemize}[leftmargin=0pt]\vspace{-0.6cm}
          \item[] Number of users performing same task
          \item[] Number of users involved in process
      \end{itemize}\\ 
     &
      Proneness to human error &\vspace{-0.6cm}
      \begin{itemize}[leftmargin=0pt]
          \item[] Number of exceptions
          \item[] Time to solve exception
      \end{itemize} \\
      \bottomrule
    \end{tabular}
    \vspace{\baselineskip}
    \caption{Process characteristics evaluation framework}
    \vspace{-0.8cm}
    \label{tab:PCEF}
\end{table}

\subsubsection{Task perspective}
The task perspective refers to the execution of process activities. 
Its criteria are standardization, maturity, determinism, and failure rate.

First, standardization refers to a process's degree of structure. %
In standardized processes, every process element is unambiguous, and the execution order remains the same in each process instance.
As a result, stakeholders receive the same outcome from a standardized process~\cite{lacity2015robotic,leshob2018towards}.
Thus, we examine the execution order and the number of process variants to measure a process's standardization.
We can, for instance, analyze predecessors and successors of the process of interest.
Ideally, the order of execution remains the same and equals to the desired process flow.
 
Maturity indicates that no frequent changes to the process flow are observable.
Therefore, processes need to be specified and predictable over a period in time \cite{beetz2019robotic,leshob2018towards}.
Mature processes usually terminate successfully and show a comparably low number of variants~\cite{syed2020robotic,huang2019applying}.
The evaluation focuses on the number of process variants and the difference between the ideal and variant process paths.

Determinism is one of the most distinctive criteria to assess the viability of RPA. 
Deterministic activities consist of logical execution steps without any form of cognitive assessment \cite{cooper2019robotic,jimenez.2019,syed2020robotic,yatskiv2019improved}.
This is a fundamental requirement for software robots since human judgment aggravates automation.
To fulfill the criterion, logical and rule-based steps suffice to describe a process.
Hence, the evaluation examines manual interactions and execution time.

Last, the failure rate relates to self loops to repair previous executions and a non-recoverable unsuccessful termination.
A low failure rate leverages automation.
The failure rate subsequently focuses on the amount of deviations from the ideal process flow caused by failures, and their respective causes \cite{beetz2019robotic,fung2014criteria,jimenez.2019,syed2020robotic,wanner2019process}.
A high failure rate might correspond to poor standardization, maturity or determinism as the causes for exceptions.

\subsubsection{Time perspective}
The criteria listed under the time perspective focus on the duration and frequency of processes and process steps.

First of all, the frequency describes the absolute number a process step occurs over time. 
The execution frequency is high when tasks are repeated daily and in high transaction volumes \cite{cooper2019robotic,fernandez2018impacts}.
The criterion measures the number of an activity's occurrences in a certain period.
\par
Additionally, the framework includes the duration which expresses the time required to execute a process or an activity.
The duration needed to execute a process or an activity is a quantifiable indicator of the time-saving potential.
\par
The final time-based criterion of the framework is urgency which describes how critical the immediate execution of a process step is.
The delayed execution may cause an increasing overall duration, or may hinder progress.
Software robots are working 24/7, unlike users with relatively short time frames.
For this reason, the evaluation focuses on the time needed to react to execute such urgent tasks.

\subsubsection{Data perspective}
In many processes, information is processed in multiple systems. 
Therefore, the data perspective resembles the structuredness of data.
If a robot shall process data, the data source must be digital~\cite{osmundsen.2019}.
Moreover, the data must at least be semi-structured to enable automation~\cite{beetz2019robotic}.
When a process involves handling data, users may perform simple operations to extract it from the source and enter it into a system~\cite{huang2019applying,kokina2019early,moffitt2018robotic,wanner2019process,zhang2019intelligent}. %
This is a crucial requirement for the successful interpretation and execution of process steps.
To evaluate this criterion the data source is analyzed.
Typically structured data is in semi-structured forms like spreadsheets, websites, or emails.
Unstructured and hardly accessible data impedes RPA.

\subsubsection{System perspective}
The fourth perspective in the framework is related to the underlying systems. The perspective poses the interaction with interfaces, and the stability of information systems.
\par
Due to our preceding research we added the criterion interface to our framework.
The criterion interface is evaluated by identifying whether the task could be solved using software robots.
Here, the time spent in an application's interface and the number of required execution steps serve as indicators.

Another system-related criterion is the stability.
Ideally, systems and applications involved in process automation are stable.
During process execution, all operations on the user interface perform accordingly, and users only seldom experience interruptions \cite{fung2014criteria,penttinen2018choose,wanner2019process}.
A stable operating system also relates to this criterion.
It guarantees the absence of system related exceptions during automation.
For analyzing system stability, we propose the number of soft- and hardware exceptions.
Important in practice is to distinguish between exceptions caused by the systems or applications themselves and external factors such as capacity errors or connection.
\par
The last system-related criterion in the framework is the number of systems.
It deals with process parts or activities that interact with multiple information systems.
Consequently, the interaction between systems is necessary, but no value is added when performed by a person \cite{Ivancic.2019,jimenez.2019,osmundsen.2019}.
In fact, robots outperform humans in atomic operations, like copy and 
paste~\cite{anagnoste2017robotic,fernandez2018impacts}.
Thus, automation candidate tasks transfer information from one to other systems. 
The potential of more involved systems is higher, if these are running stably.

\subsubsection{Human perspective}
The last perspective deals with humans computer interaction focusing on the human.
The perspective comes with two peculiarities, resources and proneness to human error.

The framework includes resources as criterion to highlight the number of users involved in the process.
Especially frequent activities require resources to deal with the volume of work.
This criteria can be assessed from two view points.
First, based on the number of users performing the same task.
Second, multiple users contribute to an activity's instance. \cite{kokina2019early,wanner2019process}.
To assess the resource savings, we utilize the count of users performing the same task, and the number of users involved in one task instance.
\par
The last aspect in the PCEF is the proneness to human errors as a criterion.
Humans tend to erroneous behavior when executing monotonous and voluminous tasks which results in such errors that solely relate to human nature
\cite{fernandez2018impacts,Hindel.2020,Ivancic.2019,jimenez.2019,santos2019toward,yatskiv2019improved}.
Eliminating such mistakes with business rules or robots yields to additional savings regarding costs and time.
Measuring the error proneness relies on the number of human mistakes and the required time to fix those.

%% file: sections/05_evaluation.tex
\section{Evaluation}

The evaluation focuses on event logs generated through PAIS.
Event logs reveal insights about the business process and its execution.
We aim at an objective evaluation by using a publicly available data-set to show the applicability of the PCEF \cite{Data.2019}.
We determine process characteristics with Process Mining Software\footnote{https://www.celonis.com}.
Hence, we test the framework for its applicability in a practical environment.

The candidate process describes a P2P process 
of a multinational coatings and paints enterprise.
Due to its administrative character, it is a suitable candidate for automation.
In this case, RPA minimizes manual work and increases efficiency at the enterprise's bottom line.
The candidate process covers the steps from creating a purchase order to the clearance of the invoice.
A purchase order contains at least one purchase order item.
An item stores attributes describing the resources involved, value of events and anonymized company information.

In total, the data set includes more than 1.5 million events, and 251,734 purchase order items (cases) in 76,349 purchase orders.
To illustrate the structure of the event log, Table \ref{tab:event log example} visualizes an event log from the data set.

\begin{table}[htb]
\centering
\caption{Exemplary event with contextual attributes from the event log}
\label{tab:event log example}
\begin{tabular}{@{}ll@{}}
\toprule
Attribute                             & Value                          \\ \midrule
Case ID                          & 2000000000\_00001              \\
Activity                         & Record Goods Receipt           \\
Resource                         & user\_000                      \\
Complete Timestamp               & 2018/03/06 07:44:00.000        \\
Variant                          & Variant 65                     \\
Variant index                    & 65                             \\
(case) Company                   & companyID\_0000                \\
(case) Document Type             & EC Purchase order              \\
(case) GR-Based Inv. Verif.      & false                          \\
(case) Goods Receipt             & true                           \\
(case) Item                      & 1                              \\
(case) Item Category             & 3-way match, invoice before GR \\
(case) Item Type                 & Standard                       \\
(case) Name                      & vendor\_0000                   \\
(case) Purch. Doc. Category name & Purchase order                 \\
(case) Purchasing Document       & 2000000000                     \\
(case) Source                    & sourceSystemID\_0000           \\
(case) Spend area text           & CAPEX \& SOCS                  \\
(case) Spend classification text & NPR                            \\
(case) Sub spend area text       & Facility Management            \\
(case) Vendor                    & vendorID\_0000                 \\
Cumulative net worth (EUR)       & 298.0                          \\
User                             & user\_000                      \\ \bottomrule
\end{tabular}
\end{table}

To analyse the framework, we focus on the paths related to \emph{Item Category: 3-way match, invoice before GR}. We further drill down selecting the most common variants (\~90\%) in 2018.
These filters result in 197,010 cases with 136 process variants.

Examining the event log reveals that traces including the manual activity 'Change Quantity' take a month longer on average.

Thus, we select 'Change Quantity' as our process step of interest, 
and apply our framework to evaluate the activity.
Note that we consider the deletion of a purchase order item and the reoccurrence of `Change Quantity' as incompliant.

\noindent\textit{Standardization.}
The criterion examines a process's degree of structure, and it relates to a low number of overall variants.

Our analysis of `Change Quantity' reveals that it has five valid predecessors covering 95\% of all incoming process paths, and two valid successors that cover 94\% of all outgoing traces.
Additionally, we examine the activity's process segment in different business units.

Every business unit conducts the activity in the same context.
Consequently, we identify a logical and structured process flow. 
The assessment shows that the process is rather standardized, since 95\% of all preceding and 93\% of all following activities are compliant and follow a certain pattern.

\noindent\textit{Maturity.}
The maturity expresses the number of compliant process variants which establish over time. In total there are 25 variants containing the activity. Out of these 25 variants, 22 are following compliant pre- and sucessors while three are incompliant. There are 2 variants reworking the activity and one which causes the deletion of purchase order items.

\noindent\textit{Determinism.}
To assess the criterion, we must know the steps done on the user interface and the respective throughput time of steps need to be evaluated.
The event log does not include information about the performance of the activity 'Change Quantity' on the presentation layer.
Therefore, the criterion can not be evaluated for this data set.

\noindent\textit{Failure Rate.}
In this process, the execution fails when a self-loop occurs or the process ends with the activity `Delete Purchase Order Item'.
Reworking `Change Quantity' occurs in 3,91\%, and the process determination with order item deletion happens in 1,42\%. 
Since we only consider the outcome of one activity, we ignore the full process context, since we cannot determine which cases actually terminated and which are still running.

The resulting failure rate of the process is 5,33\%.

\noindent\textit{Frequency.}
The average number of 'Change Quantity' occurrences is 31 times a day.
Although the execution of 'Change Quantity' varies month by month, it occurs at least 379 times a month.
Regarding frequency, the activity is a valid automation candidate.

\noindent\textit{Duration.}
The duration expresses an activity's impact on the overall process throughput time and its own required time. 
Information about its own execution time to execute is missing. 
However, while processes without have an average throughput time of 64 days, processes including the activity take 93 days on average.

\noindent\textit{Urgency.}
The majority of the tasks are executed during the main business hours.
But, `Change Quantity' quite often occurs outside of these hours.
This incidence might indicate, that certain purchase orders need fast reaction.
Automation runs all the time and minimizes the delay caused by the working hours of a user.

\noindent\textit{Structuredness of data.}
To perform the activity 'Change Quantity', workers modify the purchase order document. 
If the source data containing the new quantity and the purchase order document are structured data objects, a software robot could perform the transaction.
However, as the event log does not contain related information, 
we cannot evaluate the criterion.

\noindent\textit{Interfaces.}
This aspect analyses the number of interfaces and the interactions with these interfaces. 
The event log does not contain such information.
Thus, we cannot evaluate the criterion.

\noindent\textit{Stability.}
The stability corresponds to a low number of deviating paths and software exceptions.
The event log does not include information about exceptions and their cause.
Thus, the criterion can not be evaluated for the data set.

\noindent\textit{Number of systems.}
Since the event log originates from an SAP ECC system, which is a roofing system, we cannot determine whether there are more systems involved in the process.
Therefore, our evaluation of the number of systems is incomplete.

\noindent\textit{Resources}
Analyzing the number of users that execute the 'Change Quantity' unveils that 138 different users execute the task. 
With the successful implementation of a robot, we can spare working time of these users.

\noindent\textit{Proneness to human error.} 
Since the process step is exclusively executed by users, we assume that all related errors are of human origin. 
Since only about every twentieth case fails, we assume the process is rather stable, and software robots could not leverage better performance.

As we demonstrate, the RCEF aids in determining characteristics of processes or process steps which are automation candidates.
Although we could not assess all criteria in our case, the evaluation provides important insights.

The process flow is quite standardized.
On one hand the activity is in 22 different compliant process variants, on the other hand only three infrequent variants lead to non-compliance.
In total, the overall failure rate is 5,33\%, highlighting that 94,67\% of all executions are fully compliant. 
The activity constantly occurs during the observation, 31 times per day on average.
Since 'Change Quantity' occurs outside usual business hours, we assume the execution is urgent to a certain extent and is restrained by manual execution.
Moreover, we can spare working hours of 138 users, if we can automate the activity successfully. 
Without being able to assess the determinism, we cannot assess the viability of RPA implementation for the activity. 
Still, without knowing anything about the process context, the framework enables wide assessment.

However, the application of the framework also revealed some deficits when validating the efficacy and validity through process mining software.
First, missing attributes such as starting timestamps in the event log impede the possibility to assess typically easy to evaluate criteria like the execution time or execution urgency.
Second, the possible lack of information about exceptions in the event log inhibits the ability to distinguish between a  system-related stability or human error caused issue.
Third, crucial information about the interaction on the user interface is missing and prevents the examination of the criteria determinism, structuredness of data, interfaces and number of systems.
The missing information prevents the extraction of information such as the degree of deterministic behavior when executing a sequence of steps, the throughput time for individual steps or the number of applications and web-based systems used.
To extend the detail of information, the use of an user interaction logger \cite{bosco2019discovering} can bridge the gap between front-end and back-end information gathering.

%% file: sections/06_discussion.tex
\section{Conclusion}

By conducting a literature review, this study identified process activity characteristics for RPA.
These insights were used to develop a process characteristic evaluation framework that assesses the suitability of process activities for RPA.
The framework includes a set of thirteen criteria grouped into five evaluation perspectives, enabling the examination of a process activity on different reference levels.
This abstraction of a process step emphasizes its connections to preceding and succeeding steps and provides a concrete decision support considering the most important factors involved.

Therefore, this study offers practitioners a guideline to evaluate a process activity for an RPA implementation effort through the application of process mining.
The analysis reveals the standardization of the activity, its maturity over time, the determinism of execution steps, the failure rate, the volume of executions with respect to completion and reaction times, the structuredness of data used, the interaction on the user interface, stability and number of systems, users involved and the cause of exceptions related to human error.
This study proved the efficacy and validity of the framework by evaluating a process activity through event logs out of a real-life data set.
Based on the universal perspectives within the framework, the applicability in different organizations and industries is seen as given.

Despite the demonstration and application of the framework, it is tested only with one data set and process.
The evaluation has shown that not all criteria can be tested against this data set and to guarantee generalization, the framework must be validated through application to multiple and different kinds of processes. In particular, the framework contains qualitative criteria that could not be tested with the data set. Further evaluation of these criteria -- e.g. through case studies -- is necessary.
Another important aspect is that the data set was anonymized and modified before publishing, limiting the accessible information stored in the event logs.
Further, the assumptions made about the data set, including filters set for the focus on one execution flow, limit the significance of the evaluation results.
Additionally, the criteria must be tested for redundancy and their respective evaluation examples need further validation and extension.

Although these factors impair the evaluation of the framework, they offer various opportunities for future research.
First, the framework should be evaluated in different ways to ensure comprehensive validation.
These can include the application of the framework to new data sets as well as the assessment of a process with a process owner.
Conducting expert interviews to assess the usefulness of the framework is another option to account for the solution objective.
Changing the evaluation approach and substituting process mining through robotic process mining \cite{bosco2019discovering} can also widen the scope of information extraction.
Second, the increasing number of articles on this topic generates new insights that can derive additional perspectives and criteria.
By conducting a case study research further evaluation examples could surface and help practitioners to examine their processes. 
Finally, possible advances also include the quantification \cite{wanner2019process} or the weighting of criteria to signal if the process activity is suitable for RPA or not.

%% file: main.bbl
\begin{thebibliography}{10}
\providecommand{\url}[1]{\texttt{#1}}
\providecommand{\urlprefix}{URL }
\providecommand{\doi}[1]{https://doi.org/#1}

\bibitem{van2018robotic}
van~der Aalst, W.M., Bichler, M., Heinzl, A.: Robotic process automation.
  Business \& Information Systems Engineering  \textbf{60},  269--272 (2018)

\bibitem{aguirre2017automation}
Aguirre, S., Rodriguez, A.: Automation of a business process using robotic
  process automation (rpa): A case study. In: Workshop on Engineering
  Applications. pp. 65--71. Springer (2017)

\bibitem{anagnoste2017robotic}
Anagnoste, S.: Robotic automation process - the next major revolution in terms
  of back office operations improvement. Proceedings of the International
  Conference on Business Excellence  \textbf{11}(1),  676 -- 686 (2017)

\bibitem{asatiani2016turning}
Asatiani, A., Penttinen, E.: Turning robotic process automation into commercial
  success--case opuscapita. Journal of Information Technology Teaching Cases
  \textbf{6}(2),  67--74 (2016)

\bibitem{beetz2019robotic}
Beetz, R., Riedl, Y.: Robotic process automation: Developing a multi-criteria
  evaluation model for the selection of automatable business processes. In:
  AMCIS2019. AIS Electronic Library (2019)

\bibitem{bosco2019discovering}
Bosco, A., Augusto, A., Dumas, M., La~Rosa, M., Fortino, G.: Discovering
  automatable routines from user interaction logs. In: International Conference
  on Business Process Management. pp. 144--162. Springer (2019)

\bibitem{cooper2019robotic}
Cooper, L.A., Holderness~Jr, D.K., Sorensen, T.L., Wood, D.A.: Robotic process
  automation in public accounting. Accounting Horizons  \textbf{33}(4),  15--35
  (2019)

\bibitem{Data.2019}
van Dongen, B.: Bpi challenge 2019 (2019),
  \url{https://doi.org/10.4121/uuid:d06aff4b-79f0-45e6-8ec8-e19730c248f1},
  4TU.Centre for Research Data. Dataset

\bibitem{fernandez2018impacts}
Fernandez, D., Aman, A.: Impacts of robotic process automation on global
  accounting services. Asian Journal of Accounting and Governance  \textbf{9},
  123--132 (2018)

\bibitem{fung2014criteria}
Fung, H.P.: Criteria, use cases and effects of information technology process
  automation (itpa). Advances in Robotics \& Automation  \textbf{3} (2014)

\bibitem{geyer2018process}
Geyer-Klingeberg, J., Nakladal, J., Baldauf, F., Veit, F., van~der Aalst, W.,
  Casati, F., Conforti, R., de~Leoni, M., Dumas, M.: Process mining and robotic
  process automation: A perfect match. In: BPM (Dissertation/Demos/Industry).
  pp. 124--131 (2018)

\bibitem{Hallikainen.2018}
Hallikainen, P., Bekkhus, R., Pan, S.L.: How opuscapita used internal rpa
  capabilities to offer services to clients. MIS Quarterly Executive
  \textbf{17},  41--52 (01 2018)

\bibitem{Hindel.2020}
Hindel, J., Cabrera~Pérez, L., Stierle, M.: {Robotic} {Process} {Automation}:
  {Hype} or {Hope}? In: Proceedings of the 15th International Conference on
  Wirtschaftsinformatik (2020). \doi{10.30844/wi{\_}2020{\_}r6-hindel}

\bibitem{Hofmann-Rpa}
Hofmann, P., Samp, C., Urbach, N.: Robotic process automation. Electronic
  Markets  \textbf{30},  99–106 (08 2019). \doi{10.1007/s12525-019-00365-8}

\bibitem{huang2019applying}
Huang, F., Vasarhelyi, M.A.: Applying robotic process automation (rpa) in
  auditing: A framework. International Journal of Accounting Information
  Systems  \textbf{35},  100433 (2019)

\bibitem{Ivancic.2019}
Ivan{\v{c}}i{\'c}, L., Vugec, D.S., Vuk{\v{s}}i{\'c}, V.B.: Robotic process
  automation: Systematic literature review. In: International Conference on
  Business Process Management. pp. 280--295. Springer (2019)

\bibitem{jimenez.2019}
Jimenez-Ramirez, A., Reijers, H.A., Barba, I., Del~Valle, C.: A method to
  improve the early stages of the robotic process automation lifecycle. In:
  International Conference on Advanced Information Systems Engineering. pp.
  446--461. Springer (2019)

\bibitem{kaya2019impact}
Kaya, C.T., Turkyilmaz, M., Birol, B.: Impact of rpa technologies on accounting
  systems. Journal of Accounting \& Finance  \textbf{82},  235--250 (2019)

\bibitem{kokina2019early}
Kokina, J., Blanchette, S.: Early evidence of digital labor in accounting:
  Innovation with robotic process automation. International Journal of
  Accounting Information Systems  \textbf{35},  100431 (2019)

\bibitem{lacity2015robotic}
Lacity, M.C., Willcocks, L.P.: Robotic process automation at telefónica o2.
  MIS Quarterly Executive  \textbf{15}(1),  21--35 (2016)

\bibitem{lacity2017new}
Lacity, M.C., Willcocks, L.P.: A new approach to automating services. MIT Sloan
  Management Review  (2017)

\bibitem{leshob2018towards}
Leshob, A., Bourgouin, A., Renard, L.: Towards a process analysis approach to
  adopt robotic process automation. In: 2018 IEEE 15th International Conference
  on e-Business Engineering (ICEBE). pp. 46--53. IEEE (2018)

\bibitem{mendling2018machine}
Mendling, J., Decker, G., Hull, R., Reijers, H.A., Weber, I.: How do machine
  learning, robotic process automation, and blockchains affect the human factor
  in business process management? Communications of the Association for
  Information Systems  \textbf{43}(1), ~19 (2018)

\bibitem{moffitt2018robotic}
Moffitt, K.C., Rozario, A.M., Vasarhelyi, M.A.: Robotic process automation for
  auditing. Journal of Emerging Technologies in Accounting  \textbf{15}(1),
  1--10 (2018)

\bibitem{osmundsen.2019}
Osmundsen, K., Iden, J., Bygstad, B.: Organizing robotic process automation:
  Balancing loose and tight coupling. In: Proceedings of the 52nd Hawaii
  International Conference on System Sciences (2019)

\bibitem{penttinen2018choose}
Penttinen, E., Kasslin, H., Asatiani, A.: How to choose between robotic process
  automation and back-end system automation? In: European Conference on
  Information Systems 2018 (2018)

\bibitem{ratia2018robotic}
Ratia, M., Myll{\"a}rniemi, J., Helander, N.: Robotic process
  automation-creating value by digitalizing work in the private healthcare? In:
  Proceedings of the 22nd International Academic Mindtrek Conference. pp.
  222--227 (2018)

\bibitem{romao2019robotic}
Romao, M., Costa, J., Costa, C.J.: Robotic process automation: A case study in
  the banking industry. In: 2019 14th Iberian Conference on Information Systems
  and Technologies (CISTI). pp.~1--6. IEEE (2019)

\bibitem{santos2019toward}
Santos, F., Pereira, R., Vasconcelos, J.B.: Toward robotic process automation
  implementation: an end-to-end perspective. Business Process Management
  Journal  \textbf{26}(2),  405--420 (2019)

\bibitem{seasongood2016not}
Seasongood, S.: Not just for the assembly line: A case for robotics in
  accounting and finance. Financial Executive  \textbf{32}(1),  31--32 (2016)

\bibitem{sonmez2019conceptual}
S\"{o}nmez, O.E., B\"{o}rek{\c{c}}i, D.Y.: A conceptual study on rpas as of
  intelligent automation. In: International Conference on Intelligent and Fuzzy
  Systems. pp. 65--72. Springer (2019)

\bibitem{syed2020robotic}
Syed, R., Suriadi, S., Adams, M., Bandara, W., Leemans, S.J., Ouyang, C., ter
  Hofstede, A.H., van~de Weerd, I., Wynn, M.T., Reijers, H.A.: Robotic process
  automation: Contemporary themes and challenges. Computers in Industry
  \textbf{115},  103162 (2020)

\bibitem{wanner2019process}
Wanner, J., Hofmann, A., Fischer, M., Imgrund, F., Janiesch, C.,
  Geyer-Klingeberg, J.: Process selection in rpa projects--towards a
  quantifiable method of decision making. In: ICIS 2019 Proceedings (2019)

\bibitem{webster.2002}
Webster, J., Watson, R.T.: Analyzing the past to prepare for the future:
  Writing a literature review. MIS Quarterly  \textbf{26}(2),  xiii--xxiii
  (2002)

\bibitem{willcocks2016service}
Willcocks, L.P., Lacity, M.: Service automation robots and the future of work.
  SB Publishing (2016)

\bibitem{willcocks2015robotic}
Willcocks, L.P., Lacity, M., Craig, A.: Robotic process automation at
  xchanging. The Outsourcing Unit Working Research Paper Series
  \textbf{15}(03) (2015)

\bibitem{yatskiv2019improved}
Yatskiv, S., Voytyuk, I., Yatskiv, N., Kushnir, O., Trufanova, Y., Panasyuk,
  V.: Improved method of software automation testing based on the robotic
  process automation technology. In: 2019 9th International Conference on
  Advanced Computer Information Technologies (ACIT). pp. 293--296. IEEE (2019)

\bibitem{zhang2019intelligent}
Zhang, C.: Intelligent process automation in audit. Journal of Emerging
  Technologies in Accounting  \textbf{16}(2),  69--88 (2019)

\end{thebibliography}
